\documentclass[manuscript,screen]{acmart}
\acmJournal{TOSEM}
\renewcommand\footnotetextcopyrightpermission[1]{}

\setcopyright{none}

\usepackage{etoolbox}
\usepackage{amsmath}
\usepackage{amsfonts}
\usepackage{multicol}
\usepackage{latexsym}
\usepackage{multirow}
\usepackage{graphicx}
\usepackage{pifont}
\usepackage{url}
\usepackage{hyperref}
\usepackage{amsthm}
\usepackage[linesnumbered,commentsnumbered,ruled,vlined]{algorithm2e}

\SetKwRepeat{Do}{do}{while}

\usepackage{array}
\newcolumntype{L}[1]{>{\raggedright\let\newline\\\arraybackslash\hspace{0pt}}m{#1}}
\newcolumntype{C}[1]{>{\centering\let\newline\\\arraybackslash\hspace{0pt}}m{#1}}
\newcolumntype{R}[1]{>{\raggedleft\let\newline\\\arraybackslash\hspace{0pt}}m{#1}}

\usepackage{tcolorbox}
\definecolor{promptbg}{RGB}{244,248,252}
\definecolor{promptframe}{RGB}{94,121,148}

\newtcolorbox{takeaway}{
    colback=promptbg,
    colframe=promptframe,
    boxrule=0.6pt,
    arc=3pt,
    left=8pt, right=8pt, top=8pt, bottom=8pt,
    boxsep=0pt
}

\newcommand{\ToolName}{\textsc{ConCovUp}}

\usepackage{semantic}
\usepackage[utf8]{inputenc}
\usepackage{cleveref}
\crefname{section}{§}{§§}
\Crefname{section}{§}{§§}
\usepackage{wrapfig}

\usepackage{balance,booktabs,multirow,bm,wrapfig,graphicx,caption,subcaption,mathtools,dashbox,pifont,xcolor,colortbl,color,soul,amsfonts,enumitem,float}

\title[ConCovUp: Effective Agent-Based Test Driver Generation for Concurrency Testing]{ConCovUp: Effective Agent-Based Test Driver Generation \\ for Concurrency Testing}

\usepackage{mdframed} 
\mdfdefinestyle{promptbox}{
    backgroundcolor=promptbg,
    linecolor=promptframe,
    linewidth=0.6pt,
    roundcorner=3pt,
    innertopmargin=8pt,
    innerbottommargin=8pt,
    innerleftmargin=9pt,
    innerrightmargin=9pt
}

\makeatletter
\newcommand{\@authornotemarknum}[1]{%
  \g@addto@macro\@currentauthors{\footnotemark[#1]}%
}
\newcommand{\authornotemark}[1][]{%
  \ifx\relax#1\relax
    \g@addto@macro\addresses{\@authornotemark}%
  \else
    \g@addto@macro\addresses{\@authornotemarknum{#1}}%
  \fi
}
\makeatother

\begin{document}

 \author{Yuandao Cai}
\authornote{Both authors contributed equally to this research.}
 \affiliation{%
 	\institution{Independent Researcher}
 	\city{Hong Kong}
 	\country{China}
 }
 \email{ycaibb@cse.ust.hk}

  \author{Shuhao Fu}
\authornotemark[1]
 \affiliation{%
 	\institution{The Hong Kong University of Science and Technology}
 	\city{Hong Kong}
 	\country{China}
 }
 \email{sfual@cse.ust.hk}

 \author{Wensheng Tang}
 \affiliation{%
 	\institution{Independent Researcher}
 	\city{Hong Kong}
 	\country{China}
 }
 \email{wtangae@cse.ust.hk}

 \author{Cheng Wen}
 \affiliation{%
 	\institution{Xidian University}
 	\city{Xi'an}
 	\country{China}
 }
 \email{wencheng@xidian.edu.cn}

 \author{Shengchao Qin}
 \affiliation{%
 	\institution{Xidian University}
 	\city{Xi'an}
 	\country{China}
 }
 \email{shengchao.qin@gmail.com}

 \author{Charles Zhang}
 \affiliation{%
 	\institution{The Hong Kong University of Science and Technology}
 	\city{Hong Kong}
 	\country{China}
 }
 \email{charlesz@cse.ust.hk}

 \renewcommand{\shortauthors}{Yuandao Cai, Shuhao Fu, et al.}

\begin{abstract}

Concurrency testing is essential to improve the reliability and security of multi-threaded programs. 
Dynamic analysis tools (e.g., TSan) depend on high-quality test drivers that reach critical shared-memory interactions at runtime. However, current testing practices predominantly focus on sequential logic, leaving a critical gap in automated concurrent test generation.
Recently, Large Language Models (LLMs) have shown promise in generating sequential tests, but they struggle to produce effective concurrent tests without a deep understanding of concurrency semantics.
This paper presents \ToolName, a novel multi-agent framework that combines LLMs with program analysis. \ToolName\ grounds test generation in static analysis to extract shared memory accesses and their calling contexts. To trigger hard-to-reach accesses, it introduces an LLM-driven backward tracing approach, leveraging the model's semantic reasoning to deduce concrete inputs that satisfy complex path constraints, and iteratively refines the generated tests via dynamic execution feedback.
Our evaluation on nine real-world C/C++ libraries shows that \ToolName\ improves average Shared Memory Access Pair Coverage (SMAP Coverage) from 36.6\% to 68.1\% over the general Claude Code agent baseline.

\end{abstract}

\keywords{Concurrent Test Generation,  Large Language Models, Coverage}

\maketitle

\section{Introduction}

\sloppy
Concurrent programming is ubiquitous for achieving high performance in modern software systems, yet it introduces complex, non-deterministic behaviors that are difficult to exercise systematically~\cite{cai-icse25,cai-icse24,cai-pldi21,cai-fse22}. Unfortunately, current software testing practices remain heavily skewed towards sequential logic. This gap is particularly obvious in open-source libraries: although these components are widely assumed to be thread-safe, they often lack the necessary multi-threaded test drivers to validate this assumption~\cite{SamakRJ15-pdli15,SamakTR16-oopsla16,gong-sosp23}. 
Developers typically rely on sequential unit tests, which, while sufficient for functional validation, fail to exercise internal synchronization interactions. Consequently, without test drivers to orchestrate concurrent invocations, the thread safety of these libraries remains under-tested.

Researchers have developed seminal dynamic analysis techniques~\cite{li-oopsla25, period-icse2022, jeff-oopsl16, jeff-pldi14,congreybox-fuzz-asplos24}, such as predictive trace analysis~\cite{cai-icse23, zhang-pldi25} and dynamic data race detection~\cite{fastrack-pldi09, mosaad-popl23}. However, the effectiveness of these tools is limited by the quality of the input test cases. If the provided test driver does not actively spawn threads to create interactions on shared memory, these dynamic tools receive little useful concurrent execution signal. Therefore, the primary challenge lies in automatically generating effective concurrent test drivers, a difficult task without prior knowledge of the concurrency semantics required to manipulate shared objects across threads.

\textbf{Limitations of Existing Works.}
Previous works~\cite{AutoConTest-icse16,SamakRJ15-pdli15,SamakR14-oopsla14,SamakR15-fse15} typically generate concurrent tests by leveraging existing sequential tests or assembling method call sequences to identify possible shared object interactions. However, these approaches are inherently limited by the quality and diversity of the underlying sequential tests. If the sequential tests lack sufficient coverage, these tools cannot synthesize effective concurrent scenarios. 
Furthermore, these methods lack a holistic semantic understanding of the program, often embedded within both documentation and the source code itself (e.g., inline comments and identifier names)~\cite{zheng-icse25, chow-issta23} or complex code logic across data flows~\cite{shi-pldi18}, which is necessary to infer intricate state setups or essential concurrent invocations.

\textbf{Limitations of LLMs and General Coding Agents.}
While LLMs have advanced sequential test generation~\cite{coverup-fse25, jin-ase24, liu-issta25}, applying general coding agents to real-world concurrent testing faces two key challenges.
\textit{First, LLMs struggle to comprehend global concurrency semantics.} Concurrency-relevant behaviors often stem from complex interactions across multiple files and modules, necessitating a repository-level understanding of shared memory accesses. However, LLMs are constrained by finite context windows, preventing them from ingesting entire repositories. Furthermore, because LLMs are trained primarily on sequential text, their inherent linear reasoning struggles to capture the non-deterministic nature of thread interleavings.
\textit{Second, existing LLM-based coding agents (e.g., Claude Code and OpenCode) lack the specialized skills or tools required to analyze concurrency.} Standard code search or retrieval mechanisms cannot provide the deep semantic insights needed to identify shared variables and complex call chains. As our experiments demonstrate, without concurrency-aware tooling, general coding agents exhibit a weak ability to generate effective concurrent tests. Consequently, how to architect an agentic workflow for concurrent test generation remains a critical open question.

\textbf{Our Agent-Based Generative Approach.}
To address these key challenges, we introduce \ToolName, a novel multi-agent framework designed for automated concurrent test generation. At a high level, \ToolName\ combines the reasoning capabilities of Large Language Models (LLMs) with program analysis to connect natural language comprehension with structural code semantics. By equipping our agents with specialized static analysis tools, \ToolName\ identifies the program locations of shared memory accesses and extracts their calling contexts. Furthermore, by identifying feasible execution paths to these targeted accesses, \ToolName\ guides the LLMs to generate concurrent tests for hard-to-reach shared-memory interactions. This approach reduces the need to ingest entire repositories into limited context windows while preserving repository-level concurrency context.

\begin{figure*}[t]
    \centering
    \includegraphics[width=1\textwidth]{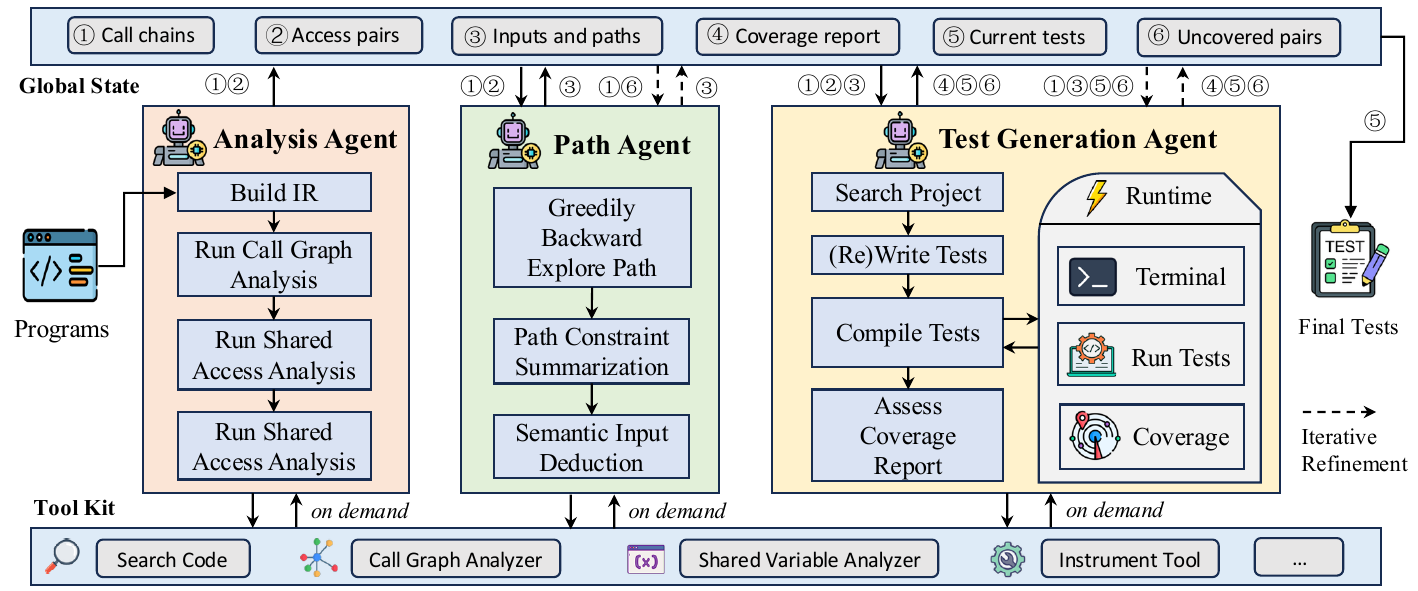}
	   \caption{Overview of \ToolName's multi-agent workflow: the Analysis Agent identifies static target pairs, the Path Agent derives path and input constraints, and the Test Generation Agent refines concurrent test drivers with coverage feedback.}
 \label{fig:conconvup-workflow}
\end{figure*}

Specifically, Figure~\ref{fig:conconvup-workflow} illustrates the architecture of \ToolName, which operates in three key stages.
\textit{First}, to fill the tooling gap of existing coding agents, an \textbf{Analysis Agent} employs whole-program static analysis to extract shared variables, shared memory accesses, and their calling contexts, which are structural elements that LLMs natively struggle to retrieve using standard search mechanisms.
\textit{Second}, to resolve the intricate state setups required for concurrent invocations, a \textbf{Path Agent} relies on a static Control Flow Graph (CFG) to perform a heuristic-guided backward greedy search. This search extracts feasible execution paths while intentionally avoiding opaque environmental APIs (e.g., I/O) and complex data structures. The agent then summarizes the path constraints using a hybrid representation of natural language and code snippets. Instead of relying on heavyweight external SMT solvers~\cite{luo2026agentic}, the agent leverages the LLM's deep semantic comprehension to directly deduce the concrete variable values and initial states required to satisfy these constraints.
\textit{Third}, leveraging the extracted semantic contexts and the deduced concrete inputs, a \textbf{Test Generation Agent} synthesizes multi-threaded test drivers. Crucially, because the LLM's input deduction is heuristic, \ToolName\ operates within a continuous feedback loop: it monitors the dynamic runtime coverage of the targeted shared memory accesses. If certain accesses remain uncovered or if the generated inputs fail to compile/execute, these specific failures are fed back to the Path Agent to explore alternative execution paths and refine its semantic reasoning, thereby improving coverage of hard-to-reach concurrent behaviors.

Our evaluation on 9 real-world C/C++ libraries totaling about 1,000 kLoC shows that \ToolName\ improves average Shared Memory Access Pair Coverage (SMAP Coverage) from 36.6\% to 68.1\% over the general Claude Code agent baseline.
The ablation study further shows that static target identification alone provides only limited gains, while the full workflow, which combines target analysis, backward path reasoning, and iterative feedback, raises coverage from 39.2\% to 68.1\%.
We also find that model capability materially affects the result: Claude Sonnet 4.6 achieves the strongest average SMAP Coverage, compared with 55.9\% for GPT 5.4 and 28.1\% for Kimi K2.5.

In summary, this paper makes the following contributions:

\begin{itemize}
\item We address key limitations of applying LLMs and coding agents to concurrent test generation by proposing a novel agent-based framework that combines LLMs with program analysis.
\item We introduce a Path Agent with a backward tracing mechanism that combines CFG-based heuristic search and semantic constraint satisfaction, empowering the LLM to derive concrete inputs for uncovered targets without the overhead of formal constraint solvers.
\item We implement and evaluate \ToolName\ on 9 real-world C/C++ libraries totaling approximately 1,000 kLoC, demonstrating clear improvements in SMAP Coverage compared to a general Claude Code agent baseline.
\end{itemize}

\textbf{Paper Organization.}
The rest of the paper is organized as follows.
Section~\ref{sec:motivation} positions \ToolName\ against existing test-generation and symbolic-execution approaches and presents a motivating example.
Section~\ref{sec:pre} defines the shared-memory access-pair abstraction and SMAP Coverage.
Section~\ref{appro} describes the three-agent workflow of \ToolName, and Section~\ref{sec:imple} summarizes its LLVM-based static analysis and coverage instrumentation.
Section~\ref{sec:eval-setup} presents the experimental setup, followed by the effectiveness, ablation, model-sensitivity, and case-study results.
Finally, Section~\ref{sec:threats} discusses threats to validity and future work, and Section~\ref{sec:conclusion} concludes the paper.

\section{\ToolName\ in a Nutshell}
\label{sec:motivation}

In this section,
we first discuss the related work.
We then highlight the essence of \ToolName\ for effective concurrent test generation.

\subsection{Related Work and Their Limitations}

Many recent works focus on generating sequential tests using LLMs. 
The first paradigm for context augmentation aims to address the limitations of LLMs' context windows and their lack of global code understanding. For instance, ChatUniTest~\cite{chatunittest-fse24} proposes an Adaptive Focal Context mechanism to precisely tailor prompts, while ASTER~\cite{aster-icse25} introduces a rigorous static analysis pipeline to extract structured information, thereby guiding LLMs to generate high-coverage test code that aligns with human developers' practices.
The second paradigm with intention and semantic guidance focuses on mitigating LLMs' shortcomings in handling complex mock behaviors and generating precise assertions. For example, IntUT~\cite{nan-icse25} injects explicit test intentions (e.g., test inputs, mock behaviors, and expected outcomes) into LLMs, improving branch coverage in industrial-level Java projects.
The third paradigm uses execution feedback and agentic refinement, which emphasizes embedding LLMs into a closed-loop "generation-validation-repair" workflow, utilizing compiler errors or dynamic runtime information for self-correction~\cite{agonetest-ase24, coverup-fse25}. 
However, all these works focus exclusively on sequential logic, ignoring the capability to reason about concurrency and shared memory accesses,
which distinguishes them from our work.

\textbf{Concurrent Test Generation.}
Some previous works focus on generating concurrent codes~\cite{AutoConTest-icse16,SamakRJ15-pdli15,SamakR14-oopsla14,SamakR15-fse15}.
Specifically, Narada~\cite{SamakRJ15-pdli15, SamakR14-oopsla14} is a trace-based approach, which relies on recording execution traces of existing sequential tests, identifying objects that could be shared, and constructing a concurrent test harness using those captured object references.
If the sequential test suite does not contain a specific object state required to trigger a complex race, Narada cannot produce the necessary setup code. It is bound by the quality and coverage of the seed suite.
Similar to previous works~\cite{SamakRJ15-pdli15, AutoConTest-icse16}, \ToolName\ allows the LLM to extract crucial context from repository artifacts, when available. By analyzing source code comments, documentation, and sequential tests, the agent can improve the quality of the generated concurrent tests.
However, a key distinction is that \ToolName\ does not depend on the presence of sequential tests. Instead, it leverages the reasoning capabilities of LLMs, guided by static concurrency analysis, to autonomously synthesize concurrent tests from scratch.

Finally, unlike techniques focused on exploring the vast space of thread schedules, such as stateless model checking~\cite{jeff-pldi15}, controlled concurrency testing~\cite{period-icse2022, learned-cct-oopsal20}, and predictive analysis~\cite{cai-icse23, cai-icse24, cai-icse25}, our work does not aim to maximize interleaving diversity directly. Instead, our approach is orthogonal to these methods. 
By generating semantically valid drivers that reach critical synchronization points, \ToolName\ provides the necessary entry points for fine-grained schedule exploration tools to operate effectively.

\textbf{Neural-Enhanced Symbolic Execution.}
Recent advancements have explored the integration of neural models into symbolic execution frameworks~\cite{klee-osdi08} to address traditional limitations. For instance, \textit{Agentic Concolic Execution}~\cite{luo2026agentic} leverages LLMs to assist forward concolic execution, primarily aiming to enhance overall code coverage.
In contrast, \ToolName\ employs a \textit{directed backward execution} strategy: our approach anchors its search from specific shared memory accesses and works backwards for synthesizing targeted concurrent tests.
Another notable approach is \textit{Neuro-Symbolic Execution}~\cite{ndss2019neuro}, which trains a dedicated neural network to approximate complex constraints and accelerate the underlying SMT solving process. Unlike their approach, which still relies on mathematical constraint solving, \ToolName\ does not rely on heavyweight SMT solvers (e.g., Z3). Instead of accumulating complex mathematical formulas that often lead to high computational overhead and scalability walls, our methodology operates at a higher semantic abstraction level. \ToolName\ summarizes path conditions into natural language constraints and directly leverages the LLM's reasoning capabilities to deduce concrete inputs. This design reduces dependence on solver-heavy path exploration for complex data structures and opaque library calls.

\subsection{Our Multi-Agent Generative Approach}

To the best of our knowledge, \ToolName\ is the first agent-based approach to generate test drivers for concurrency testing. We employ Shared Memory Access Pair Coverage (SMAP Coverage)~\cite{wen-ase14, hong-issta12} as our evaluation metric, which assesses whether conflicting access pairs, such as write-write and write-read operations executed by different threads on a given shared variable, are adequately covered by the test cases. Prior work has shown that these inter-thread data dependencies are important execution targets for studying atomicity and order violations~\cite{shan-oopsla10}.

\begin{figure*}[t]
    \centering
    \includegraphics[width=1\textwidth]{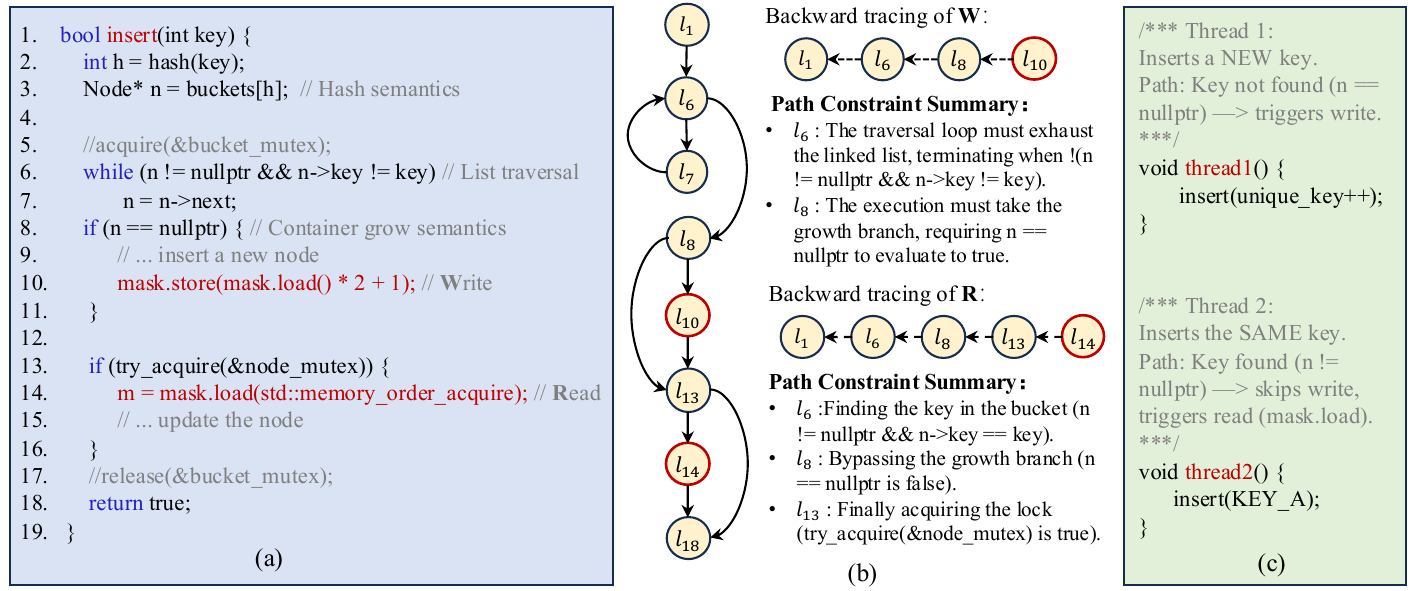}
   \caption{Motivating example of SMAP Coverage-guided concurrent test generation: (a) shows a thread-unsafe function with an atomicity violation, while (b) and (c) illustrate backward tracing and test generation.}
 \label{fig:moti-example}
\end{figure*}

We illustrate how \ToolName\ operates using the concurrent insert function of a hash map, shown in~Figure~\ref{fig:moti-example} (a).
Its control flow graph is shown in~Figure~\ref{fig:moti-example} (b) with $l$ denoting the line.
In this snippet, an unsafe interaction may arise if one thread resizes the container (writing to the global variable $mask$ at line 10) while another thread concurrently reads the $mask$ (line 14) without proper synchronization. To exercise this potential atomicity interaction, a test driver need spawn two threads that concurrently execute the exact Write-Read pair on the shared variable $mask$.
Relying solely on a general coding agent to generate such a test is unreliable, especially for a large concurrent codebase. 
The agent often fails to comprehend the complex control flow (e.g., hash semantics, linked-list traversals) required to reach these specific lines, resulting in tests that execute shallow paths and miss the deep concurrent interactions. \ToolName\ addresses this through its three-stage multi-agent workflow:

\textbf{1. Static Analysis (Analysis Agent):}
Instead of feeding the entire codebase to the LLM blindly, the Analysis Agent first performs whole-program static analysis. It automatically identifies $mask$ as a shared variable and extracts the target Write-Read pair: the write operation $mask.store(...)$ at line 10, and the read operation $mask.load(...)$ at line 14.

\textbf{2. Backward Tracing \& Input Deduction (Path Agent):}
To reach these specific accesses, the Path Agent performs a backward search on the CFG to extract the necessary path conditions, summarizing them into natural language constraints.
For Thread 1 to reach the write at line 10, the Path Agent generates the NL constraint summary as shown in in~Figure~\ref{fig:moti-example} (b).
For Thread 2 to reach the read at line 14 (while skipping the write), the NL constraint summary is also shown in~Figure~\ref{fig:moti-example} (b).
Crucially, traditional symbolic execution and SMT solvers struggle with such constraints due to the opaque hash functions (line 2) and unbounded heap-based list traversals (lines 6-7). Instead of relying on heavyweight solvers, \ToolName\ leverages the LLM's deep semantic comprehension. The LLM interprets the summaries and helps deduce that satisfying the first path requires inserting a \textit{new, unique key}, whereas the second path requires inserting an \textit{already existing key}.

\textbf{3. Test Generation (Test Generation Agent):} Equipped with the deduced concrete inputs, the Test Generation Agent synthesizes the final executable test driver. It constructs the necessary concurrent function invocations to trigger the exact target interleaving, as illustrated in Figure~\ref{fig:moti-example}(c).
By assigning the deduced inputs to specific threads, \ToolName\ steers execution along hard-to-reach paths and exercises the targeted Write-Read conflict. The primary goal of our approach is to synthesize test drivers that improve coverage of these shared memory access pairs; runtime checkers such as ThreadSanitizer can be applied separately as auxiliary analyses.

\section{Preliminaries}
\label{sec:pre}
This section defines the key concepts, metrics, and notations used throughout the paper. 

\smallskip
\textbf{Basic Notions.}
Let \( V_{\mathrm{shared}} \) be the set of shared variables in a target program that can be accessed concurrently across multiple threads (e.g., global variables, objects escaping thread boundaries). 
For a given shared variable \( v \in V_{\mathrm{shared}} \), we define an \textit{access point} as a specific source code location (file and line number) where a single read or write operation occurs. Each access point is assigned a unique identifier. The set of all access points for \( v \) is denoted as \( A_v \), and the union of all shared accesses across the program is defined as \( A_{\mathrm{shared}} = \bigcup_{v \in V_{\mathrm{shared}}} A_v \).
Furthermore, let \( F_{\mathrm{root}} \) denote the set of public entry functions (i.e., API functions) that, when invoked concurrently, may trigger one or more shared accesses in \( A_{\mathrm{shared}} \).

\smallskip
\textbf{Shared Memory Access Pair Coverage.}
To measure the effectiveness of the generated concurrent tests, we formalize the metric of \textit{Shared Memory Access Pair Coverage}. 
Let a runtime access operation be a tuple \( a = \langle \tau, op, l, addr \rangle \), where \( \tau \) denotes the thread identifier, \( op \in \{\text{Write}, \text{Read}\} \) represents the operation type, \( l \in A_{\mathrm{shared}} \) is the context-insensitive access point corresponding to a shared variable, and \( addr \) is the dynamic memory address accessed at runtime.
We define a static target access pair as \( p = (l_i, l_j) \), where \( l_i, l_j \in A_{\mathrm{shared}} \) satisfy the following conditions: (1) \textbf{Shared Target:} \( l_i \) and \( l_j \) access the same underlying shared variable \( v \). (2) \textbf{Conflict:} at least one of the two locations is a write access, effectively capturing W-W, W-R, and R-W interactions.
Let \( P_{\mathrm{total}} \) be the set of static target access pairs identified in the target program. A pair \( p = (l_i, l_j) \) is covered if the execution contains runtime accesses \( a_i \) and \( a_j \) such that \( a_i.l = l_i \), \( a_j.l = l_j \), \( a_i.\tau \neq a_j.\tau \), and \( a_i.addr = a_j.addr \).
Let \( P_{\mathrm{covered}} \subseteq P_{\mathrm{total}} \) be the subset of pairs covered during the dynamic execution of the tests. The coverage is formally defined as \( |P_{\mathrm{covered}}| / |P_{\mathrm{total}}| \). We refer to this metric as SMAP Coverage in the rest of the paper. The execution results of generated tests naturally partition the target pairs into covered and uncovered sets, directly guiding the subsequent refinement.

Note that our coverage metric is evaluated \textit{context-insensitively}. Specifically, we consider a target pair covered as long as its constituent static access points are triggered by different threads on the same dynamic memory address, regardless of the specific function call contexts (i.e., call stacks) through which these accesses are reached.
We further discuss this in the Section~\ref{sec:threats}.

\smallskip
\textbf{Goal.}
Given a target C/C++ open-source program, \ToolName\ aims to automatically synthesize \emph{concurrent test drivers}, which are multi-threaded programs that invoke functions in \( F_{\mathrm{root}} \) to maximize SMAP Coverage. By maximizing this coverage, \ToolName\ exercises concurrency-relevant shared-memory interactions and provides stronger executions for downstream dynamic analyses.

\smallskip
\textbf{Iteration Budgets.}
Static analysis is over-approximated, possibly identifying shared memory accesses that are semantically unreachable (dead code) or guarded by unsatisfiable constraints. Without a termination condition, the LLM agent could waste infinite cycles attempting to reach these infeasible targets. To prevent resource exhaustion, \ToolName\ employs a configurable refinement budget, \( K_{\mathrm{refine}} \). This bounds the dynamic feedback loop to \( K_{\mathrm{refine}} \) iterations per test, ensuring the process terminates even if some statically identified targets remain uncovered.
We set the \( K_{\mathrm{refine}} \) to three in our experiments.

\section{\ToolName\ in Detail}
\label{appro}

\ToolName\ is a multi-agent framework designed to automatically generate test drivers for concurrency testing by integrating static analysis with LLM agents. The workflow of \ToolName\ is orchestrated across three distinct agentic phases.

First, the Analysis Agent uses static analysis tools to comprehensively analyze the target program. 
Initially, it invokes a Call Graph Analyzer to pinpoint the top-level entry functions and extract the precise call graphs leading to each shared memory access.
Subsequently, it employs a Shared Access Analyzer to identify shared variables and their corresponding memory access points.

The second phase focuses on the reachability of each memory access within an identified pair. Here, the Path Agent performs backward tracing and semantic input deduction. Instead of accumulating complex mathematical formulas for a traditional SMT solver, the agent summarizes the path conditions into natural language constraints. Relying on its code comprehension capabilities, the LLM heuristically explores the execution paths and deduces the concrete inputs required to satisfy these constraints. This prompt-based reasoning reduces dependence on heavyweight formal solvers and their overhead on complex data structures and loops. The agent then formulates these deduced inputs and environment setup requirements, feeding them to the Test Generation Agent.

In the third phase, guided by the flagged shared variables, their access points, and the concrete inputs deduced in the previous stage, the Test Generation Agent synthesizes the initial batch of concurrent test drivers. These generated tests are then compiled and executed using a custom-built instrumentation tool to dynamically evaluate execution coverage.
Crucially, the system operates on an iterative feedback loop between Phase 2 and Phase 3. Based on the coverage feedback from the custom instrumentation tool, if a target access pair remains uncovered, the system loops back: the Path Agent re-evaluates the semantic constraints or explores alternative execution paths, and the Test Generation Agent refines the test drivers accordingly. This collaborative process iterates until either all target access pairs are successfully covered or a predefined iteration limit is reached.

\subsection{Analysis Agent with Static Analysis Tools}

\begin{figure*}[t]
    \centering
    \includegraphics[width=1\textwidth]{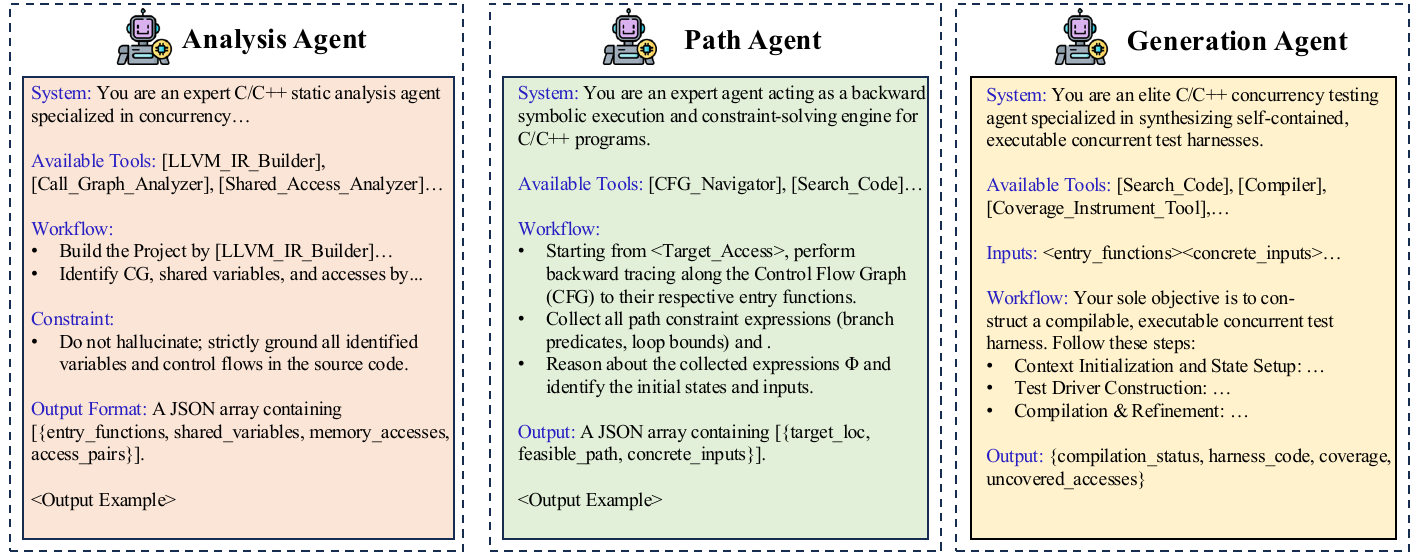}
   \caption{Prompt structure of \ToolName's multi-agent workflow, showing the task roles, tool interfaces, and structured outputs used by each agent.}
 \label{fig:prompts}
\end{figure*}

The Analysis Agent orchestrates a suite of static analysis tools to analyze the target source code. Its primary objective is to extract potential shared memory accesses and identify appropriate entry functions, thereby providing the targets for the subsequent reachability analysis and guiding the test generation process. \ToolName\ begins by addressing a key challenge in analyzing open programs (e.g., libraries): identifying possible shared memory accesses without a full application execution context.
To achieve this autonomously, the Analysis Agent is initialized with a specialized system prompt that defines its persona, available toolset, and strict output constraints. The core prompt structure is outlined in~Figure~\ref{fig:prompts}.

\smallskip
\textbf{Call Graph Analyzer:}
\ToolName\ prioritizes public entry points over internal functions, as invoking these entry points exercises the underlying internal logic. The agent invokes the \texttt{[Call\_Graph\_Analyzer]} to construct a global Call Graph (CG) of the target program. By traversing the CG, the framework identifies the top-level API functions, defined as the Root Functions (\( F_{\mathrm{root}} \)). These root functions serve as the primary entry points for the multi-threaded test drivers because generated tests invoke them to trigger the deeply embedded shared accesses.

\smallskip
\textbf{Shared Access Analyzer:}
Identifying shared memory accesses in open programs is difficult; traditional flow-sensitive analyses often struggle with inter-procedural aliasing when the calling context from an upper-layer application is absent. To address this, the agent utilizes the \texttt{[Shared\_Access\_Analyzer]}, which employs a flow-insensitive yet field-sensitive pointer analysis~\cite{jeff-icse16}. 
The flow-insensitivity over-approximates the control flow, allowing the system to conservatively model the possible function invocations and thread interleavings. To counterbalance this over-approximation and reduce false positives, field-sensitivity is applied to prune infeasible aliasing relationships.

A variable is classified into the shared variable set (\( V_{\mathrm{shared}} \)) if it satisfies any of the following criteria: 
\begin{enumerate}
    \item It is passed as a pointer or reference parameter to a Root Function (\( f \in F_{\mathrm{root}} \)).
    \item It is protected by a synchronization primitive, such as a lock (encompassing both heap-allocated memory and global variables).
    \item It is passed as an argument to a thread creation API (e.g., \texttt{pthread\_create}).
\end{enumerate}

Once \( V_{\mathrm{shared}} \) is established, the analyzer scans the codebase to extract the set of corresponding memory access points (\( A_{\mathrm{shared}} \)). To prepare for the dynamic testing phases, the Analysis Agent groups these individual context-insensitive access locations (e.g., \( l_i, l_j \in A_{\mathrm{shared}} \)) into the set of static target access pairs (\( P_{\mathrm{total}} \)). The agent ensures that for each pair \( (l_i, l_j) \in P_{\mathrm{total}} \), both locations access the same underlying shared variable \( v \), and at least one operation is a write access (capturing W-W, W-R, and R-W conflicts). By mapping the call chains from the Root Functions (\( F_{\mathrm{root}} \)) down to these context-insensitive access pairs, the Analysis Agent establishes the concurrency semantics required to output the final JSON array, directly guiding the following phases.

\begin{example}
To illustrate this process, consider Figure~\ref{fig:moti-example}. The Analysis Agent first identifies \texttt{insert} as a Root Function (\( f \in F_{\mathrm{root}} \)). During the shared access analysis, the agent observes the global variable $mask$. Because $mask$ is accessed within a critical section protected by $node\_mutex$ (via $try\_acquire$ at line 13), it satisfies the second criterion and is added to the shared variable set (\( V_{\mathrm{shared}} \)). Subsequently, the agent extracts its memory access points: a write operation at line 10 ($mask.store(...)$) and a read operation at line 14 ($mask.load(...)$) inside the locked region. These two code locations, denoted as \( l_{10} \) and \( l_{14} \), are then grouped into a valid access pair \( (l_{10}, l_{14}) \in P_{\mathrm{total}} \) representing a Write-Read conflict. Finally, the agent outputs this pair along with its call chain in the JSON array, setting the stage for the reachability analysis in Phase II.
\end{example}

\subsection{Path Agent for Semantic Path Exploration}
\label{sec:backward_agent}

To generate the precise input values required to reach the constituent accesses of each target pair, \ToolName\ introduces the Path Agent. Operating as the second phase of our framework, this agent initially analyzes all identified shared memory accesses to provide baseline execution paths and concrete inputs for the Test Generation Agent (Phase III). 
In addition, it also has an iterative refinement capability. When Phase III reports (via dynamic coverage feedback) that certain deeply embedded accesses remain uncovered---often due to complex control-flow dependencies or rigid environmental requirements---the Path Agent is re-invoked. During this feedback loop, it re-analyzes the execution paths, deprioritizes previously failed routes, and synthesizes alternative concrete inputs.

To achieve this autonomously, the Path Agent is driven by a specialized prompt (as detailed in Figure~\ref{fig:prompts}). The prompt defines its persona as an expert backward execution and constraint-solving engine for C/C++ programs. It equips the agent with navigation and search tools, such as \texttt{[CFG\_Navigator]} and \texttt{[Search\_Code]}, and dictates a strict workflow: starting from a constituent target access location, the agent performs backward tracing along the CFG to an entry function, collects path constraint expressions, reasons about these expressions to identify initial states, and finally outputs a JSON array containing the target location, feasible path, and concrete inputs.

Unlike traditional dynamic symbolic execution (DSE) engines, which frequently suffer from path explosion and struggle with complex data structures, this agent operates at a higher semantic abstraction level. It is designed to autonomously discover feasible execution paths, summarize path constraints, and perform semantic input deduction through LLMs. Ultimately, the insights derived by this agent form a collaborative, closed-loop system with Phase III, capable of synthesizing precise concurrent tests.

\smallskip
\textbf{Backward Path Extraction.}
To reach a target memory access, the agent identifies a feasible execution route originating from a top-level entry function (Root Function). Relying on a pre-computed static CFG and its \texttt{[CFG\_Navigator]} tool, the agent performs a backward traversal starting from each constituent target access location. 

To mitigate the well-known path explosion problem inherent in path exploration, the agent employs a heuristic-guided greedy search strategy. Rather than exhaustively exploring all predecessors, the agent evaluates and prioritizes paths based on a defined reachability cost. Specifically, this heuristic favors paths with fewer complex branch conditions and actively penalizes paths that involve heavy environmental interactions, such as system calls, network I/O, or complex file operations. By intentionally avoiding these opaque, environment-dependent APIs, the agent focuses its computational effort on paths driven purely by internal program logic and input parameters. During iterative refinement, if a previously selected path fails to achieve coverage in Phase III, the agent dynamically adjusts its search space, pruning the failed branch and selecting the next optimal path from its priority queue.

\smallskip
\textbf{Path Summarization and Semantic Input Deduction.}
Once a promising path is identified, the agent transitions to constraint extraction and semantic input deduction. Traditional DSE engines~\cite{klee-osdi08, klee-fse20} mechanically translate every instruction into low-level bitvector formulas, an approach that frequently overwhelms formal solvers. In contrast, our Path Agent leverages the LLM's native reasoning capabilities to formulate and resolve path constraints at a semantic level. The agent collects branch predicates, loop bounds, and state transformations along the path, summarizing them into a high-level representation that captures the core program requirements and variable relationships without the instruction-level translation. We denote the collected path constraints as \( \Phi \).

Instead of relying on SMT solvers, the LLM acts as an autonomous reasoning engine. It analyzes the collected constraint expressions \( \Phi \) and directly deduces the initial states and concrete inputs required to satisfy them based on its semantic understanding of the code. \ToolName\ incorporates a robust backtracking mechanism during this deduction phase: if the LLM determines that the current path constraints are logically conflicting or unsatisfiable (UNSAT)---indicating that the path is logically dead or blocked by mutually exclusive conditions---the agent discards the path, backtracks to the extraction phase, and selects the next most promising route.
When a path is considered satisfiable, the agent reasons the concrete variable values required to satisfy the branch conditions. Finally, the agent passes both the identified execution path and the concrete input values forward to the Test Generation Agent. With this semantic guidance, Phase III can synthesize a targeted test driver that navigates the complex control flow to trigger the concurrent memory access.

\begin{example}
    Returning to Figure~\ref{fig:moti-example}, the Path Agent deduces inputs to trigger the Write-Read conflict on $mask$. For Thread 1 to reach the write operation at line 10, the agent traces backward and summarizes the constraints: the traversal loop at line 6 exhausts the list (\texttt{n == nullptr}), and the execution takes the growth branch at line 8. Conversely, for Thread 2 to reach the read at line 14 while skipping the write, the constraints dictate that the loop at line 6 finds the key (\texttt{n != nullptr \&\& n->key == key}), skips the growth branch at line 8, and successfully acquires the lock at line 13. 
\end{example}

\subsection{Generation Agent for Concurrent Tests}
\label{sec:test_generation_agent}

The Concurrent Test Generation Agent serves as the synthesis core of \ToolName\ (Phase III). Guided by the static analysis targets and target entry functions extracted in Phase I and the feasible paths, path constraints, and concrete inputs provided by the Path Agent in Phase II, this phase harnesses the generative capabilities of LLMs to automatically construct executable, multi-threaded test drivers. To manage the complexity of concurrent scenarios, the agent is driven by a system prompt that defines its persona as an \textit{``elite C/C++ concurrency testing agent specialized in synthesizing self-contained, executable concurrent test harnesses.''} Its workflow is organized into four operational stages: Task Partitioning, Tool-Augmented Context Resolution, Prompt-Driven Synthesis, and Coverage Validation.

\smallskip
\textbf{Task Partitioning and Code Annotation.}
A naive approach to test generation would overwhelm the LLM by feeding it the entire set of shared variables simultaneously, leading to context dilution. To prevent this, \ToolName\ employs a \textit{Write-Anchor Partitioning Strategy}. For a given shared variable \( v \), the system fixes a specific write access \( l_i \) as the ``anchor'' and pairs it with another target access \( l_j \) (read or write). Let \( E(l) \in F_{\mathrm{root}} \) denote the selected root function whose call chain reaches access location \( l \). The task scheduler dictates the construction of a two-thread test driver: Thread 1 invokes the top-level entry function \( E(l_i) \), while Thread 2 concurrently invokes \( E(l_j) \). To further assist the LLM's attention, the source code snippets are dynamically annotated with inline comments and line-number markers directly adjacent to the target memory accesses.

\smallskip
\textbf{Tool-Augmented Context Resolution.}
A problem in generating compilable C/C++ test drivers is the \textit{Missing Context Problem}. The agent receives \texttt{<entry\_functions>} and static targets from Phase I, as well as \texttt{<concrete\_inputs>} and path constraints from Phase II, but it still has to properly initialize complex parameters and global states. To resolve this, the prompt equips the agent with \texttt{Available Tools}, notably \texttt{[Search\_Code]}. When the agent encounters an unknown custom data type or an opaque function signature, it autonomously invokes this tool to query the codebase. It dynamically retrieves source code, struct definitions, and existing sequential tests, ensuring the setup code correctly maps Phase II's semantic inputs into valid C/C++ initialization code.

\smallskip
\textbf{Prompt-Driven Test Synthesis.}
With the context resolved, the agent proceeds to the core synthesis step, strictly following the \texttt{Workflow} defined in its prompt. Because Phase II has provided candidate resolutions for the path constraints, Phase III is dedicated to the precise engineering task of code construction. The prompt instructs the agent to follow a step-by-step process:
\begin{enumerate}
    \item \textbf{Context Initialization and State Setup:} Guided by the retrieved variable, function, and structure definitions, the agent generates the code required to instantiate complex data structures, allocate memory, and set up global states.
    \item \textbf{Test Driver Construction:} To improve compilability and encourage correct thread interactions, the agent encapsulates the initialized environment, spawns the two interacting threads (e.g., using \texttt{std::thread}), invokes the target entry functions with the synthesized inputs, and cleanly joins the threads. A provided one-shot exemplar ensures a rigid, compilable output format.
\end{enumerate}

\smallskip
\textbf{Compilation, Execution, and Refinement.}
\ToolName\ operates in a closed-loop environment. The prompt instructs the agent to perform the compilation and refinement using the provided \texttt{[Compiler]} and \texttt{[Coverage\_Instrument\_Tool]}. Once compiled, \ToolName\ automatically executes the harness. A target pair is marked as covered when the runtime coverage log shows that the generated concurrent harness triggered both designated static locations and that the corresponding accesses refer to the \textit{exact same dynamic memory address}.

Finally, the agent formats its results into the requested \texttt{Output} format: a structured JSON containing \texttt{\{compilation\_status, harness\_code, coverage, uncovered\_accesses\}}. If the generated test successfully triggers the target access pair, the pair is marked as covered. For pairs that fail to trigger---often due to rigid environmental checks or precise timing requirements---the execution trace is logged in \texttt{uncovered\_accesses} and fed back to the Path Agent (Phase II). This triggers the iterative refinement process, where Phase II computes alternative paths to guide Phase III in subsequent generation attempts.

\begin{example}
    In the Figure~\ref{fig:moti-example}, based on the Phase II results, it assigns a dynamically incrementing \texttt{unique\_key++} to \texttt{thread1} to satisfy the ``new key'' precondition, and a specific \texttt{KEY\_A} (pre-initialized in the state setup step) to \texttt{thread2} to satisfy the ``existing key'' precondition. Crucially, following the concurrent scaffolding rules defined in the prompt, the agent completes the test case by explicitly spawning and joining two concurrent threads (e.g., \texttt{std::thread t1(thread1); std::thread t2(thread2); t1.join(); t2.join();}). This creates a concurrent execution in which the two functions can overlap, enabling the generated harness to trigger the targeted Write-Read interaction in the shared memory.
\end{example}

\section{Implementation}
\label{sec:imple}

\ToolName\ is implemented as a set of tool-augmented subagents in Claude Code 2.1.90 for C/C++ libraries. 
The subagents follow the workflow in Figure~\ref{fig:conconvup-workflow} and invoke specialized tools on demand, including a shared access analyzer and a coverage instrumentation tool. 
These tools provide structural program facts that cannot be reliably inferred through textual code search alone.

\emph{Static analysis tools.}
The static analysis tools are built on LLVM 15. 
The Call Graph Analyzer computes reachability from the root functions to internal functions and memory access locations. 
The Shared Access Analyzer follows the criteria used by the Analysis Agent: it conservatively marks pointer or reference arguments of root functions, objects protected by synchronization primitives, and objects passed to thread-creation APIs as candidate shared memory objects. 
It then checks reachable reads and writes for each candidate and groups conflicting accesses into access pairs. 
The analysis is field-sensitive, so different fields of the same aggregate object are represented as distinct memory locations.

\emph{Coverage instrumentation.}
\ToolName\ measures SMAP Coverage through bitcode-level instrumentation. 
For each statically identified access, the instrumentation logs the access ID, access type (read or write), executing thread ID, and dynamic memory address. 
\ToolName\ reconstructs observed pairs from this log by matching statically identified target accesses that execute in different threads on the same dynamic address, with at least one access being a write. 
The instrumentation provides coverage feedback to the agents for subsequent test refinement.

\section{Evaluation}

We evaluate our work through investigating the below questions:

\begin{itemize}
    \item (\textbf{RQ1: Overall Effectiveness}) To what degree does \ToolName\ enhance SMAP Coverage and improve concurrent test generation effectiveness compared with a general Claude Code agent baseline?
    \item (\textbf{RQ2: Ablation Study}) What are the individual and combined contributions of stage-level agents and iterative generation to the SMAP Coverage achieved by \ToolName?
    \item (\textbf{RQ3: Model Sensitivity}) How does the selection of the underlying LLM backend affect the SMAP Coverage attained by \ToolName?
\end{itemize}

\subsection{Experimental Setup}
\label{sec:eval-setup}

\textbf{Libraries.}
We select nine widely used C/C++ libraries that cover common domains in systems software, as summarized in Table~\ref{tab:subjects}.
The libraries include networking (c-ares, curl, libuv), data structures and compression (cJSON, concurrentqueue, zlib), cryptography (libsodium), and logging (spdlog, zlog), totaling approximately 1,000 kLoC.
These libraries expose public APIs and maintain shared state through global objects, heap objects, synchronization-protected data, or callback-driven execution.
This mix lets us evaluate whether generated concurrent drivers can handle different API styles and shared-memory access patterns rather than a single library family.

\begin{table}[t]
\centering
\caption{Benchmark Libraries Used in the Evaluation.}
\label{tab:subjects}
\begin{tabular}{llc}
\toprule
Library & Domain / Functionality & kLoC \\
\midrule
c-ares & Asynchronous DNS resolver & 66.9 \\
cJSON & JSON parser/generator & 24.4 \\
concurrentqueue & Lock-free queue & 327.0 \\
curl & URL transfer library & 294.9 \\
libsodium & Cryptographic library & 76.3 \\
libuv & Event-loop and asynchronous I/O & 111.8 \\
spdlog & C++ logging library & 33.7 \\
zlib & Compression library & 44.4 \\
zlog & C logging library & 19.7 \\
\bottomrule
\end{tabular}
\end{table}

\textbf{Configurations.}
We compare three configurations that share the same iterative generation budget but differ in the program-analysis support available to the agent.
\textit{General CC Agent} is our baseline. It runs in Claude Code with Claude Sonnet 4.6 and receives natural-language task instructions and the basic testing workflow, with the simplified skill shown in Appendix~\ref{app:simplified-skills}. It can use the ordinary coding-agent environment, but it is not given \ToolName's specialized static-analysis tools, backward tracing capability, or SMAP Coverage instrumentation support.
\textit{Analysis-guided} is an ablation that adds the Analysis Agent and its static tools, so the generator receives shared-memory targets and entry functions before producing tests; it does not include the Path Agent's backward tracing.
The full \ToolName\ configuration includes the Analysis Agent, the Path Agent, the Test Generation Agent, and coverage feedback.
Appendix~\ref{app:simplified-skills} summarizes the simplified skills used for the three configurations.
All configurations use \(K_{\mathrm{refine}}=3\), so the stage ablation isolates the contribution of agent-provided program facts and semantic path guidance rather than the presence of iteration itself.

\textbf{Metric.}
We use SMAP Coverage, defined in Section~\ref{sec:pre}, as the primary metric.
For each library and configuration, we aggregate SMAP Coverage over the generated test drivers that compile and execute successfully.
We also record generation time, token usage, and TSan warnings for the full \ToolName\ configuration.

\textbf{Environment and LLM Selection.}
All experiments run on Ubuntu 20.04 with an Intel(R) Xeon(R) Platinum 8358 CPU and 256GB RAM.
We configure \ToolName\ to use Claude Sonnet 4.6 as the default LLM because \ToolName\ is implemented and orchestrated in Claude Code.
For the model comparison experiments in RQ3, we also evaluate GPT 5.4 as another closed-source frontier model and Kimi K2.5 as an open-source coding-oriented model with a strong verified entry on the Terminal-Bench 2.0 leaderboard at the time of access~\cite{terminalbench20,kimi-k25}.
The comparison is intended to cover different model providers and coding-agent backends.

\subsection{RQ1: Effectiveness of \ToolName}
\label{sec:rq1}

RQ1 compares the full \ToolName\ workflow with \textit{General CC Agent}. Both configurations run in Claude Code with Claude Sonnet 4.6 and use the same iteration budget; the difference is that \ToolName\ provides static analysis, backward tracing, and coverage feedback. We use SMAP Coverage~\cite{wen-ase14, hong-issta12} as the metric.

\begin{figure}[t]
\centering
\includegraphics[width=0.95\textwidth]{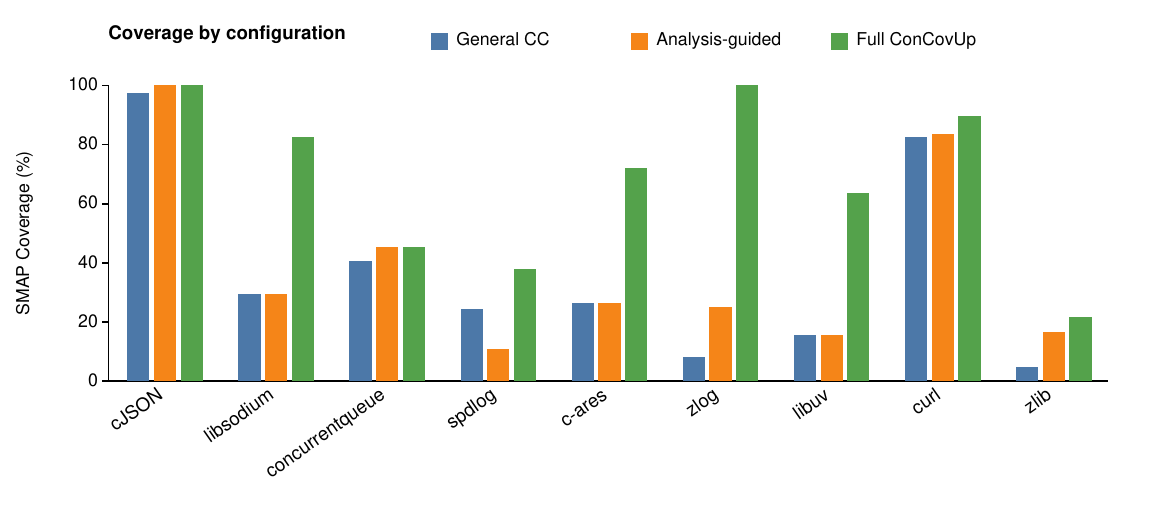}
\caption{SMAP Coverage of \ToolName\ and General/Ablated Configurations.}
\label{fig:coverage-ablation}
\end{figure}

\begin{table*}[t]
\centering
\caption{Static Target Scale, Generation Time, Token Usage, and TSan Warnings for the Full \ToolName\ Configuration.}
\label{tab:performance}
\small
\begin{tabular}{lcccccc}
\toprule
Library & \# Shared Var & \# Shared Access & Static Time & Wall Time & In / Out Tokens (k) & TSan Warnings \\
\midrule
cJSON            & 6 & 24 & 6s & 20m 8s & 333 / 43 & 5 \\
libsodium        & 11 & 26 & 7m 18s & 47m 1s & 779 / 77 & 0 \\
concurrentqueue  & 52 & 211 & 16s & 21m 39s & 454 / 39 & 0 \\
spdlog           & 17 & 43 & 24s & 52m 40s & 1043 / 104 & 0 \\
c-ares           & 24 & 130 & 24s & 51m 43s & 931 / 115 & 1 \\
zlog             & 7 & 32 & 21s & 20m 39s & 317 / 31 & 7 \\
libuv            & 21 & 59 & 1m 24s & 1h 19m & 1112 / 161 & 4\\
curl             & 7 & 56 & 2m 9s & 37m 37s & 698 / 74 & 0 \\
zlib             & 2 & 36 & 26s & 47m 35s & 696 / 105 & 2 \\
\bottomrule
\end{tabular}
\end{table*}

\smallskip
\noindent\textbf{Results.}
Figure~\ref{fig:coverage-ablation} shows that the full \ToolName\ configuration achieves 68.1\% average SMAP Coverage, compared with 36.6\% for \textit{General CC Agent}. This is a 31.5 percentage-point improvement, and the full workflow improves coverage on every evaluated library.

The largest gains appear on zlog, libsodium, libuv, and c-ares. These libraries require the generator to identify relevant shared-memory targets, choose entry functions, and construct initialization states or path conditions that expose the target pairs. The general baseline receives the testing goal and a basic workflow, but it does not receive the static targets or path constraints that \ToolName\ provides. This explains why the gap is larger on libraries where simply invoking public APIs is not enough to reach the intended shared accesses.

\ToolName\ also reaches 100.0\% SMAP Coverage on cJSON and zlog, and obtains high coverage on curl, libsodium, and c-ares. These results show that the same workflow applies across different API styles, including compact C libraries, logging libraries with global state, and networking libraries with callback- or state-driven execution.

The lower-coverage cases are also explainable. zlib requires threads to share the same \texttt{z\_stream} or related state object while respecting its API lifecycle. spdlog is header-only and template-heavy, so generated tests can execute logging APIs while leaving many target pairs less visible to instrumentation. concurrentqueue uses lock-free data-structure patterns where coverage depends on object sharing and specific interleavings. These cases leave less room for purely syntactic test construction and motivate the path- and feedback-guided parts of \ToolName.

Table~\ref{tab:performance} reports the static target scale, generation time, token usage, and TSan warnings for the full configuration.
The benchmark subjects contain 2--52 shared-memory objects and 24--211 shared accesses, while the static tools remain lightweight relative to generation time: most libraries finish static analysis in under 30 seconds, and the longest static-analysis run is 7m 18s on libsodium.
End-to-end wall time ranges from 20m 8s to 1h 19m, and token usage ranges from 317k/31k to 1112k/161k input/output tokens per library.
The generated tests also execute under ThreadSanitizer, producing 19 TSan warnings across the benchmark suite.

\begin{takeaway}
\textbf{Answer to RQ1:} \ToolName\ improves average SMAP Coverage from 36.6\% to 68.1\% over the general Claude Code agent baseline. The gains are largest when static targets, path constraints, and feedback help the generator construct shared-state executions that are hard to obtain from natural-language guidance alone.
\end{takeaway}

\subsection{RQ2: Component Contribution Ablation}
\label{sec:rq2}

\begin{figure}[t]
\centering
\includegraphics[width=0.82\textwidth]{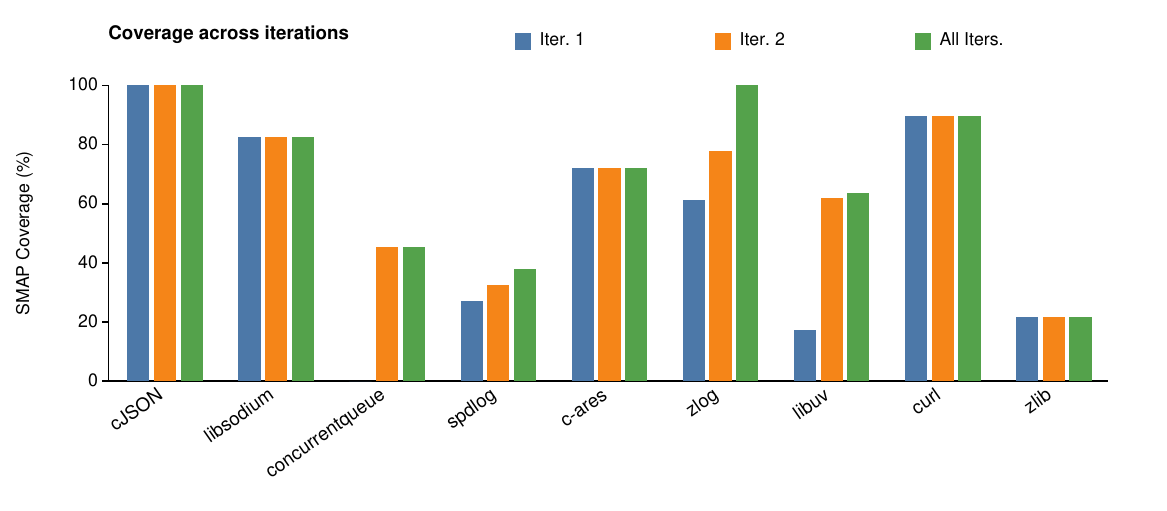}
\caption{Cumulative SMAP Coverage Gains from Iterative Test Generation.}
\label{fig:iter-percentages}
\end{figure}

RQ2 studies where the coverage gains come from. We use Figure~\ref{fig:coverage-ablation} to compare stage-level support, Figure~\ref{fig:iter-percentages} to measure how coverage accumulates over iterations, and Table~\ref{tab:iteration-counts} to report the cumulative number of generated test cases.

\begin{table}[t]
\centering
\caption{Cumulative Number of Generated Test Cases across Iterations.}
\label{tab:iteration-counts}
\begin{tabular}{lrrr}
\toprule
Project & After Iter. 1 & After Iter. 2 & After Iter. 3 \\
\midrule
c-ares          & 52 & 64 & 70 \\
cJSON           & 32 & 45 & 53 \\
concurrentqueue & 5  & 11 & 16 \\
curl            & 19 & 24 & 29 \\
libsodium       & 8  & 12 & 14 \\
libuv           & 6  & 12 & 18 \\
spdlog          & 6  & 12 & 18 \\
zlib            & 6  & 9  & 12 \\
zlog            & 18 & 40 & 51 \\
\bottomrule
\end{tabular}
\end{table}

\smallskip
\noindent\textbf{Results.}
Figure~\ref{fig:coverage-ablation} shows that \textit{Analysis-guided} achieves 39.2\% average SMAP Coverage, compared with 36.6\% for \textit{General CC Agent}. The full workflow reaches 68.1\%, a 28.9 percentage-point gain over \textit{Analysis-guided}. Figure~\ref{fig:iter-percentages} shows a related pattern for iteration: average SMAP Coverage increases from 52.4\% after the first iteration to 64.9\% after the second iteration, and then to 68.1\% after all iterations. Table~\ref{tab:iteration-counts} shows that the iterative process also expands the generated test pool, from the initial tests after Iteration 1 to the cumulative tests after Iteration 3.

The small gap between \textit{General CC Agent} and \textit{Analysis-guided} suggests that target identification alone mostly tells the generator what to test. It does not by itself solve how to reach the target paths or how to make multiple threads share the intended runtime object. This explains why Analysis-guided remains unchanged from the general baseline on libsodium, c-ares, and libuv, and why it covers fewer pairs on spdlog. In contrast, the full workflow turns static targets into path constraints, concrete setup hints, and feedback-guided repairs, which accounts for the larger gains on zlog, libsodium, libuv, and c-ares.

Iteration helps most when the first generated harness is close but misses part of the required setup. The test-count trend should be read together with Figure~\ref{fig:iter-percentages}: later iterations add more candidate tests, but coverage improves only when those tests repair missing setup, object sharing, or path reachability. Since our evaluation caps the refinement budget at \(K_{\mathrm{refine}}=3\), the reported coverage should be read as the result under this fixed budget rather than as a saturation point; additional iterations may further increase coverage when feedback leads to executable repairs for the remaining target pairs. concurrentqueue is the clearest case: coverage is 0.0\% after the first iteration and rises to 45.5\% after the second, even though the cumulative test count remains small. libuv shows a similar pattern, increasing from 17.2\% to 62.1\%, which is consistent with event-loop and callback setup being repairable after feedback. zlog improves from 61.1\% to 100.0\% as its cumulative test count grows from 18 to 40 and then 51, suggesting that iterative repair helps logging/global-state setup. By contrast, cJSON and c-ares continue to add tests, but their coverage mostly saturates after the first iteration, indicating that additional tests do not necessarily trigger new static target pairs. libsodium, curl, and zlib show a similar saturation pattern, where their covered targets are mostly determined by the initial setup that \ToolName\ synthesizes.

\begin{takeaway}
 \textbf{Answer to RQ2:} Static target information provides a small gain by identifying what to test. Most of the improvement comes from combining target information with path reasoning, concrete setup guidance, and feedback-based repair. 
\end{takeaway}

\subsection{RQ3: Impact of Different LLMs}
\label{sec:rq3}

RQ3 keeps the \ToolName\ workflow fixed and changes only the underlying LLM backend. This experiment evaluates how path reasoning, library-specific setup inference, and compilable harness synthesis vary across Claude Sonnet 4.6, GPT 5.4, and Kimi K2.5.

\begin{figure}[t]
\centering
\includegraphics[width=0.82\textwidth]{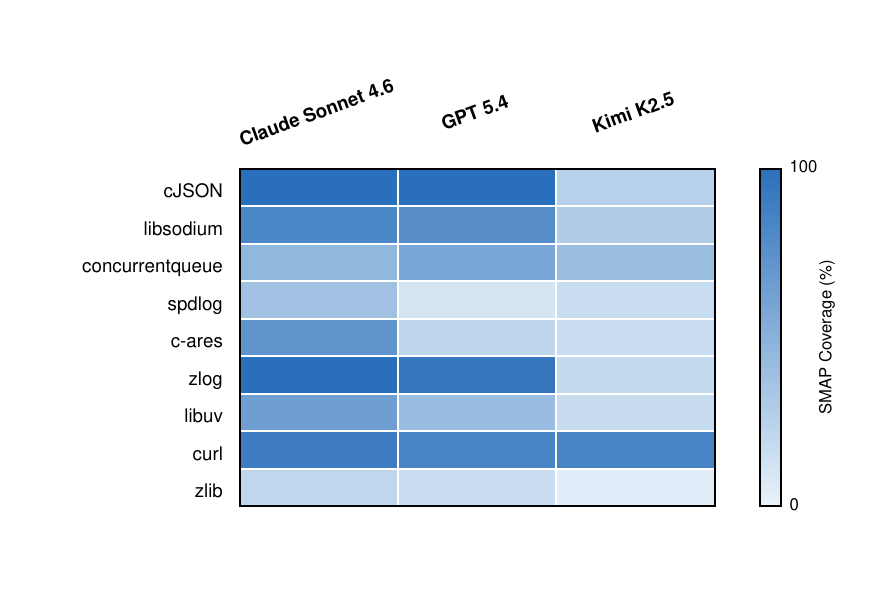}
\caption{Model sensitivity of \ToolName: darker cells indicate higher SMAP Coverage for a given library and LLM backend.}
\label{fig:model-ablation}
\end{figure}

\smallskip
\noindent\textbf{Results.}
Figure~\ref{fig:model-ablation} shows a clear model effect, with darker cells indicating higher SMAP Coverage. Claude Sonnet 4.6 achieves the highest average SMAP Coverage at 68.1\%, GPT 5.4 reaches 55.9\%, and Kimi K2.5 reaches 28.1\%. Claude Sonnet 4.6 is best or tied for best on eight of the nine libraries, which is consistent with its role as the default model in our Claude Code implementation.

The model comparison also shows that \ToolName's program-analysis workflow is useful beyond one model backend, but the quality of the generated harness still depends on the model. GPT 5.4 remains close to Claude Sonnet 4.6 on cJSON, zlog, and curl, and it outperforms Claude Sonnet 4.6 on concurrentqueue. This suggests that some data-structure scenarios can be handled well by a different model when the workflow supplies the same static targets and feedback.

The lower GPT 5.4 results on c-ares, spdlog, and zlib point to library-specific setup as a source of sensitivity. c-ares exposes hidden or internal types during setup, spdlog relies on template-heavy logging APIs, and zlib requires careful lifecycle and object sharing. Kimi K2.5 performs reasonably on curl and concurrentqueue but is lower on zlog, libuv, c-ares, and zlib. This pattern suggests that terminal or coding benchmark strength does not always translate to sustained multi-step reasoning over shared-memory targets, API states, and executable concurrent harnesses.

Overall, model choice changes the final coverage, while the workflow still provides a common structure for each backend: static targets define what to cover, path reasoning proposes how to reach it, and feedback helps repair generated harnesses.

\begin{takeaway}
\textbf{Answer to RQ3:} Model choice has a clear effect on SMAP Coverage. Claude Sonnet 4.6 provides the highest average result, GPT 5.4 remains competitive on several libraries, and Kimi K2.5 shows that coding-agent benchmark strength does not always translate into high SMAP Coverage~\cite{terminalbench20,kimi-k25}.
\end{takeaway}

\subsection{Explaining Remaining Coverage Gaps}
\label{sec:case-study-smap-coverage}

The RQ1--RQ3 results show that several lower-coverage cases are not random artifacts. To understand these cases, we reviewed the generated tests and iteration logs in the \texttt{hybrid/results\_iterative} artifact tree. The logs point to three recurring constraints: object identity across threads, the gap between diagnosis and executable repair, and instrumentation visibility.

\textbf{Object identity across threads.}
In \texttt{zlib}, generated tests call relevant APIs such as \texttt{deflate}, \texttt{deflateResetKeep}, and \texttt{deflateParams}, but workers often allocate separate \texttt{z\_stream} instances. The tests are concurrent, yet the observed accesses do not converge on the same object. A similar pattern appears in \texttt{libsodium}, where the logs report that remaining uncovered pairs access different dynamic addresses for the same access identifiers. These cases explain why API relevance alone is not enough for SMAP Coverage: the harness also has to preserve object identity across threads.

\textbf{Diagnosis versus executable repair.}
For \texttt{zlog} and \texttt{zlib}, later iterations often describe the right repair direction, such as sharing state to repair aliasing, but the generated code still allocates fresh per-thread objects or manipulates unstable internal structures. \texttt{c-ares} exposes a related buildability problem: generated tests often combine public calls with brittle internal setup and fail on hidden types such as \texttt{ares\_slist\_t}. These tests can be semantically close to the intended behavior, but compile-invalid harnesses do not improve SMAP Coverage.

\textbf{Instrumentation visibility.}
\texttt{spdlog} is header-only and template-heavy; generated tests compile and TSan warnings are absent, but the full configuration still reaches only 37.8\% SMAP Coverage. This suggests that valid logging tests can execute relevant APIs without making many target access pairs jointly observable under the instrumentation. \texttt{zlog} shows a related issue, where only part of the generated tests are instrumented in early iterations. In contrast, \texttt{libuv} shows a case where iteration helps once the setup is repairable: later iterations raise coverage after weak initial scaffolding.

Overall, the case study explains both the strengths and remaining limits of \ToolName. Static analysis and path reasoning help the agent move beyond surface-level API calls, while feedback repairs many setup issues. The remaining low-coverage cases usually involve shared object identity, library lifecycle, buildability, or observability constraints, which are reasonable for this concurrency-focused metric.

\section{Threats to Validity}
\label{sec:threats}

Currently, \ToolName\ is subject to specific scope limitations. First, the tool confines its test generation to the scope of single shared variables in isolation. 
Consequently, it does not directly target cross-variable dependencies, such as multi-variable atomicity interactions or inconsistent state updates across coupled resources. 
However, the underlying framework is designed to be generalizable. This limitation can be addressed by extending the static analysis phase to identify correlated variables, such as those frequently accessed together or protected by a common lock, and prompting the LLM to produce drivers that exercise the composite invariants.

Second, \ToolName\ employs static analysis to identify candidate shared variables and their access points. However, static analysis is inherently conservative and prone to false positives, often misclassifying thread-local data as shared or flagging dynamically-infeasible code paths. 
The imprecision can lead to wasted computational effort, as the tool may generate spurious test drivers for non-shared variables or exhaust computation resources attempting to satisfy unsatisfiable path constraints. 
These limitations could be mitigated by employing highly precise static analysis, such as context-sensitive pointer analysis, or by refining the framework to support a human-in-the-loop mechanism, allowing the agent to query developers for semantic annotations regarding the true shared variables.

Finally, \ToolName\ currently focuses on reaching access points and matching dynamic addresses, but it does not explicitly diversify the data state (i.e., the specific runtime values) present during those accesses. 
To address this limitation and enhance the concurrent test generation, we could integrate finer-grained guidance metrics, such as branch coverage, path coverage, and value-range analysis, to drive the generation of test inputs that explore a wider range of the program's state space.
We leave this direction to our future exploration.

\section{Conclusion}
\label{sec:conclusion}

This paper presents \ToolName, a novel multi-agent framework that integrates LLMs with program analysis for concurrency test generation.
\ToolName\ grounds the generation process in static analysis to identify shared memory accesses and their calling contexts, and further employs an LLM-powered backward tracing approach to reach execution paths that are difficult to exercise.
Experiments on 9 real-world libraries show that \ToolName\ improves average SMAP Coverage from 36.6\% to 68.1\% over the general Claude Code agent baseline.
These results suggest that combining program-analysis evidence with coding-agent workflows is a promising direction for concurrency testing, while leaving room for future work on richer alias reasoning, more robust harness repair, and broader support for diverse build systems.

\bibliographystyle{ACM-Reference-Format}
\bibliography{concurrency}

\appendix
\section{Simplified Skills}
\label{app:simplified-skills}

This section lists simplified skills used for each configuration.
Due to page limits, we include abbreviated skill descriptions here to illustrate the workflow structure and differences among the RQ2 settings.
Please refer to the artifacts for the full skills and prompts used in our experiments.

\subsection{General CC Agent Configuration.}
This skill describes the baseline workflow given to the \textit{General CC Agent}.
The agent receives the target library and a natural-language testing workflow, but it is not given \ToolName's specialized static-analysis, call-graph, or path-analysis tools.

\begin{mdframed}[style=promptbox]
\textbf{Role.} You are an expert C/C++ concurrency testing assistant.

\textbf{Skill Goal.} Guide the agent to generate concurrent test drivers for a target C/C++ library to maximize SMAP Coverage over the generated tests that compile and execute successfully.

\textbf{Available Inputs.} The agent is given the target library source code and its public APIs.

\textbf{Workflow.}
\begin{enumerate}
    \item \textit{Inspect the target library source code directly} and identify candidate shared memory locations, including global or static variables, shared heap objects, and shared struct fields.
    \item Group accesses to the same location into alias groups, and label each access as read or write.
    \item Select public API functions that may trigger the accesses in each alias group.
    \item Generate pthread-based concurrent test drivers that attempt to exercise write-write and write-read access pairs for those alias groups.
    \item Execute the generated tests with runtime instrumentation and compute SMAP Coverage from the observed cross-thread shared-memory accesses.
    \item Analyze uncovered pairs, unreached accesses, and compile or execution failures, then revise the tests accordingly.
    \item Repeat the generation-and-feedback loop for at most three iterations, or stop earlier if coverage reaches 100\%.
\end{enumerate}

\textbf{Constraints.} Do not modify the library source code. \textit{Do not use specialized static-analysis, call-graph, or path-analysis tools.} Coverage only counts valid shared-memory access pairs observed in successful executions.

\textbf{Expected Output.} The skill asks the agent to return the generated concurrent tests and a final summary reporting covered pairs, total pairs, and overall SMAP Coverage.
\end{mdframed}

\subsection{Analysis-guided Configuration.}
This skill corresponds to the stage ablation that gives the agent static shared-memory targets and call-graph guidance, but not the backward path reasoning used by the full \ToolName\ workflow.

\begin{mdframed}[style=promptbox]
\textbf{Role.} You are an expert C/C++ concurrency testing assistant with access to static-analysis tools.

\textbf{Skill Goal.} Guide the agent to generate concurrent test drivers for a target C/C++ library to maximize SMAP Coverage over the generated tests that compile and execute successfully.

\textbf{Available Inputs.} The agent is given the target library source code and its public APIs.

\textbf{Workflow.}
\begin{enumerate}
    \item \textit{Run the static analyzer tool} to identify shared memory locations, shared-memory accesses, and alias groups, and label each access as read or write.
    \item \textit{Use the static analyzer tool to extract a call graph from the library source code} and identify exposed entry functions that can reach the target accesses.
    \item For each alias group, select public API entry points using the call graph so that the generated tests are guided toward the target accesses.
    \item Generate pthread-based concurrent test drivers that attempt to exercise write-write and write-read access pairs for those alias groups.
    \item Execute the generated tests with runtime instrumentation and compute SMAP Coverage from the observed cross-thread shared-memory accesses.
    \item Analyze uncovered pairs, unreached accesses, and compile or execution failures, then revise the tests \textit{using the static-analysis targets and static-analysis-derived call-graph guidance}.
    \item Repeat the generation-and-feedback loop for at most three iterations, or stop earlier if coverage reaches 100\%.
\end{enumerate}

\textbf{Constraints.} Do not modify the library source code. \textit{Use the static analyzer both to identify shared-memory targets and to recover the call graph used for entry-function selection,} but rely on runtime execution to measure actual coverage.

\textbf{Expected Output.} The skill asks the agent to return the generated concurrent tests and a final summary reporting covered pairs, total pairs, and overall SMAP Coverage.
\end{mdframed}

\end{document}